\documentstyle[prb,aps,epsf]{revtex} \def\narrowtext{} \tighten \twocolumn
\input epsf.sty
 
\begin{document}
\draft
 
\title{The Temperature Evolution of the Spectral Peak in High
Temperature Superconductors}
\author{M. R. Norman$^1$, A. Kaminski$^2$, J. Mesot$^3$,
and J. C. Campuzano$^{1,2}$}
\address{
         (1) Materials Sciences Division, Argonne National Laboratory,
             Argonne, IL 60439 \\
         (2) Department of Physics, University of Illinois at Chicago,
             Chicago, IL 60607\\
         (3) Laboratory for Neutron Scattering, ETH Zurich and PSI
             Villigen, CH-5232 Villigen PSI, Switzerland\\
         }

\address{%
\begin{minipage}[t]{6.0in}
\begin{abstract}
Recent photoemission data in the high temperature cuprate superconductor
Bi2212 have been interpreted in terms of a sharp spectral peak with a
temperature independent lifetime, whose weight strongly decreases upon heating.
By a detailed analysis of the data, we are able to extract the temperature
dependence of the electron self-energy, and demonstrate that this
intepretation is misleading.  Rather, the spectral peak loses its integrity
above $T_c$ due to a large reduction in the electron lifetime.
\typeout{polish abstract}
\end{abstract}
\end{minipage}}

\maketitle
\narrowtext

Angle resolved photoemission spectroscopy (ARPES) has emerged as a powerful
tool for understanding
the electronic structure of high temperature cuprate superconductors.  This
has occurred because of its unique energy and momentum resolved nature,
and its interpretation in terms of the single particle spectral
function\cite{MOHIT}.

Perhaps the most novel behavior which has been observed is the dramatic
rearrangement of the ARPES lineshape at the $(\pi,0)$ points of the Brillouin
zone when cooling below $T_c$.  Above $T_c$, one has a single broad spectrum,
indicating incoherent states.  Below, though, a gap opens up in this incoherent
spectrum, and inside this gap, a sharp peak emerges, leading to the
well-known peak-dip-hump structure \cite{PDH} which was first
observed in tunneling spectra \cite{JOHNZ}.  This onset of coherence at $T_c$
is quite dramatic, and has obvious implications for the microscopics of
high temperature superconductors \cite{MILLIS}.

How the sharp peak appears below $T_c$, though, has recently emerged
as a controversial issue.  Previous analysis \cite{PHENOM} pointed to a
dramatic increase in the low-energy lifetime as the source of its appearance.
This agrees with interpretations of microwave \cite{BONN} and
thermal \cite{ONG} conductivity measurements.  Recently, though, this picture
has been challenged \cite{LOESER,VALLA,FENG}.  Based primarily on
new high resolution data, it has been asserted
that the spectral peak width in energy remains fixed with temperature.
Rather, its weight (the quasiparticle renormalization factor, $z$) tracks
the temperature dependence of the superconducting order parameter, and thus
vanishes as $T_c$ is approached from below.  As emphasized by
Carlson {\it et al.} \cite{CARLSON}, this is not in support of the previous
``lifetime catastrophe" scenario, and casts doubt on standard interpretations
based on reduction of the low energy scattering rate due to the opening of
the superconducting gap \cite{3DELTA,NDING,CHUB}.

This new picture has a certain appeal to it, in that as Feng {\it et al.}
show, one can draw a strong correlation between the temperature dependence
of the spectral peak weight and other quantities, such as the superfluid
density and the intensity of the magnetic resonance observed by inelastic
neutron scattering \cite{FENG}.  Similar conclusions have
been reached by Ding {\it et al.} \cite{DING}, though their analysis has
some aspects of the lifetime catastrophe scenario as well.  Moreover, since
the spectrum is completely incoherent above $T_c$, it is natural to
suppose that $z$ does indeed monotonically decrease to zero as $T_c$ is
approached from below.

In this paper, we will argue that this new picture is misleading.
Our analysis is based on a methodology we have developed where the electron
self-energy is directly extracted from ARPES data \cite{NORM99}.
From this analysis, we find that the quasiparticle residue $z$ is at best a
marginally
defined quantity at low temperatures, and becomes increasingly ill defined
as $T$ increases.  Moreover, any reasonable attempt to define a $z$ leads
to a quantity which actually {\it increases} with increasing $T$ (more
properly, the energy integrated spectral weight for binding energies less than
the dip energy is temperature independent).
This is due to a reduction in the mass renormalization as $T$
increases, similar to what is obtained from a generalized Drude analysis of
optical data \cite{PUCHKOV}.  This, coupled with the increase of the low
energy scattering rate (as also seen in optics), leads to a strong
increase in the spectral peak width.
That is, the spectral peak loses its integrity (rather than simply
disappearing) due to a ``lifetime catastrophe".

The ARPES data were taken on an optimally doped ($T_c$=90K) Bi2212 sample, with
the $\Gamma-M$ axis parallel to the photon polarization vector,
and were previously reported in another connection \cite{ADAM2}.
The measurements were carried out at the Synchrotron Radiation Center in
Madison, WI, on the U1 undulator beamline, with a Scienta SES 200 electron
analyzer having an energy resolution of 16 meV and a momentum resolution of
$0.01\AA^{-1}$.

In Fig.~1, we show data taken at the $M (\pi,0)$ point of
the Brillouin zone as a function of temperature.  The leading edge of the
spectral peak is determined by the superconducting gap, whose energy stays
fairly fixed in temperature, and persists above $T_c$ (the pseudogap).
On the trailing edge, one sees a spectral dip, whose energy also remains fixed
in temperature, and becomes filled in above $T_c$.  As these two energy scales
define the two sides of the peak, this then gives the illusion that the
peak width is independent of temperature.  That such is not the case can be
clearly seen by a closer inspection of the trailing edge of the peak.
There is no question from Fig.~1 that the trailing edge is broadening
with temperature, and this broadening is what is causing the spectral dip
to fill in above $T_c$.

The earlier attempts to quantify this behavior involved separating the spectrum
into a coherent (``peak") part and an incoherent (``hump") part.  Although
this a relatively straightforward procedure at low temperatures, where these
two features are quite well defined, this becomes increasingly difficult
as the temperature is raised and the spectral dip is filled in.  The
resulting ambiguity of what to call the coherent part, and what the
incoherent part, has led to the differing conclusions in the literature.

The obvious way to overcome this difficulty is to treat the spectral function
as a unified object.  We note that the spectral function is defined as the
imaginary part of the Greens function, that is
\begin{equation}
A = -\frac{1}{\pi}\frac{Im\Sigma}{(\omega-\epsilon-Re\Sigma)^2+(Im\Sigma)^2}
\end{equation}
where $\epsilon$ is the bare energy and $\Sigma$ the Dyson self-energy.
Therefore, the temperature
evolution of the total spectral function is determined by the temperature
evolution of the self-energy.

We now briefly review the work of Ref.~\onlinecite{NORM99}, where the
procedure for extracting $\Sigma$ from ARPES data was developed.
To determine $\Sigma$ uniquely, we need to know both $ReG$ and $ImG$.
Unfortunately, from ARPES, we only know the occupied part of $ImG$,
\begin{equation}
I({\bf k},\omega) = C_{\bf k} \sum_{\delta \bf k} \int d\omega'
A({\bf k'},\omega') f(\omega') R(\omega-\omega') + B
\end{equation}
where $I$ is the photocurrent, $C$ an intensity prefactor (proportional to 
the square of the dipole matrix element between initial and final states),
$R$ the energy resolution function, and $B$ an extrinsic background, with
the sum representing the momentum resolution.
To make progress, certain assumptions have to be made.  For instance, if we
assume the spectrum is particle-hole symmetric, then the full $ImG$
(symmetrized data) can be obtained from the above equation by the identity,
$A(\omega)f(\omega)+A(-\omega)f(-\omega)=
A(\omega)$, which holds even in the presence of resolution.  Obviously,
the particle-hole symmetry assumption is a reasonable approximation 
only on the Fermi surface.  Since the spectral
peak is dispersionless around $(\pi,0)$, we will assume particle-hole
symmetry for our purposes.

A more difficult problem is that the spectrum is a continous function of
energy, and contains an extrinsic background, $B$.  Separating out what
part of the spectrum is intrinsic, and what is due to the state near the
Fermi energy, and what is due to the tail of the main valence band, is
an unresolved issue.  For our purposes here, we will make the minimal
assumption that the entire spectrum is intrinsic, which we simply cut off
at the lowest energy data were taken at ($\omega_c$=-320 meV).  As discussed
extensively in Ref.~\onlinecite{NORM99}, varying the cut-off and background
assumptions make quantitative, rather than qualitative, differences in
the results.  This was verified in the present paper as well.

Therefore, our procedure is as follows.  The photocurrent is symmetrized about
zero energy (zero energy determined by measuring the chemical potential of a
polycrystalline gold sample in electrical contact with the sample).  The
prefactor $C$ in Eq.~2 is eliminated by invoking the condition that the
integral of
$A$ is unity over the energy range considered ($\pm \omega_c$), and $B$ is
assumed zero.  This then gives $ImG$.  $ReG$ is determined by
Kramers-Kronig transformation.  From the full $G$, $Re\Sigma$ and $Im\Sigma$
is then uniquely determined ($\epsilon$ being taken as zero).  Unlike
Ref.~\onlinecite{NORM99}, since the data were obtained from a high
resolution detector, we elected to use raw data, i.e., the noise was
not filtered out, nor the resolution deconvolved.

In Fig.~2a, $Im\Sigma$, as determined from Fig.~1, is plotted for various
temperatures.  At low temperatures and energies, it is characterized by a peak
centered at zero energy due to the
superconducting energy gap, and a ``normal" part which can be treated as
a constant plus an $\omega^2$ term.  Besides the constant term, which is
surprisingly large, this is
the expected form for the self-energy for a superfluid Fermi liquid.
A maximum in $Im\Sigma$ occurs near the energy
of the spectral dip.  Beyond this, $Im\Sigma$ has a large, nearly frequency
independent, value (the slow decay at higher energies is due to the cut-off
at -320 meV, and so should not be taken seriously).  As the temperature
is raised, the zero energy peak broadens, the constant term increases, and
the $\omega^2$ term goes away.

In Fig.~2b, the quantity $\omega-Re\Sigma$ is plotted.  The low temperature,
low energy behavior is again characteristic of a superfluid Fermi liquid.
At low energies, there is a $1/\omega$ term due to the energy gap, and
a ``normal" part which is linear in $\omega$.  As expected, the zero crossing
is near the location of the spectral peak.  Beyond this, there is a minimum
near the specral dip energy, then the data are approximately linear again,
but with a smaller slope than near the zero crossing.  As the temperature is
raised, the gap (1/$\omega$) term broadens out and the low energy linear
in $\omega$ term decreases,
paralleling the behavior discussed above for $Im\Sigma$.

To further quantify this behavior, we have found that the following form
for the self-energy gives a good description of data for low energies (smaller
than the energy of the spectral dip)
\begin{eqnarray}
\omega-Re\Sigma = Z\omega(1-\frac{\Delta_r^2}{\omega^2+\Gamma_r^2}) \nonumber \\
-Im\Sigma = c + c_{FL}\omega^2 + \frac{\Delta_i^2\Gamma_i}
{\omega^2+\Gamma_i^2}
\end{eqnarray}
A discussion of this form is in order.  The term involving $\Delta$ is simply
a broadened version of the BCS self-energy, with $\Gamma$ representing a
combination of resolution and pair lifetime effects \cite{PHENOM}.  Note that
the quantity $\Delta_i/\sqrt{Z}$ in the expression for $Im\Sigma$ is
approximately equivalent to $\Delta_r$ where $Z$ is the mass factor in the
expression for $Re\Sigma$,
and that $\Gamma_r$ and $\Gamma_i$ are essentially equivalent (these
quantities would be identical by the Kramers-Kronig relations if the
self-energy had this form for all energies).  From this
expression, one sees that the occupied quasiparticle weight is approximately
(neglecting $\Gamma$) the inverse of 2$Z$ (the other half
of the quasiparticle weight lies above the Fermi energy).  Also, if the
gap term is neglected, one sees that the effective quasiparticle width
is approximately $-Im\Sigma/Z$.  Note that $Z$ is the inverse of the
quasiparticle renormalization factor, $z$.  The above assumes, of course,
that quasiparticles exist, a matter which we will address below.

In Fig.~3, the temperature variation of these coefficients, obtained from
fitting data over an energy range of $\pm$60 meV, is shown.  The most
significant finding is that $Z$ {\it decreases} with temperature.  This
implies that if quasiparticles exist, their weight {\it increases} with
temperature.
This is in contrast with earlier analyses \cite{VALLA,FENG,DING}.
We have attempted to quantify the weight in two ways.  First, we define
$z$ by taking the inverse of the derivative of $\omega-Re\Sigma$ from
Eq.~3 evaluated at the peak maximum, $\omega_p$, and multiplying by two
(the factor of two accounting for the other peak above the Fermi energy).
Second, we input Eq.~3 into Eq.~1, defining a quantity $A_{low}$, integrate
this in energy over the {\it full}
energy range ($\pm$ 320 meV) and divide by the actual integrated weight.
These quantities ($z$ and $ratio$) are plotted in Fig.~4a, and both increase
with increasing $T$.  This conclusion is further supported by the fact that
the actual integrated weight over the energy range of the fit ($\pm$ 60 meV)
relative to the total integrated weight is essentially constant with
temperature (quantity $low$ of Fig.~4a).  For this to be true, then the
weight factor must increase to compensate for the broadening of the peak.

Therefore, rather than the
peak decreasing in weight, it disappears by broadening strongly in energy.  This
can be seen directly by inspecting Fig.~2, in that as the temperature
increases, $Im\Sigma$ (Fig.~2a) in the vicinity of the peak increases in
magnitude with $T$, and $Z$ (roughly the slope in Fig.~2b near the zero
crossing) decreases with $T$.  In fact, it is the strong variation of $c$ and
$Z$, and the fact that they operate in concert, which is responsible for
the rapid variation in the effective width of the peak with $T$.
This can be quantified by two estimates of the width of the peak which are
analogous to the weight estimates discussed above,
first from the quantity $-zIm\Sigma(\omega_p)$, second from
the full width half maximum (FWHM) of $A_{low}$.
These quantities ($FW1$ and $FW2$) are plotted in
Fig.~4b.  Note, the FWHM ($FW2$) would be even larger above $T_c$ if it
were not for the pseudogap splitting the peak in two.

The above analysis is important in that it shows how coherence is lost in
the system.
It is apparent from Figs.~1 and 2 that once a temperature is reached where the
peak is no longer discernable in the data, the difference in behavior
of the self-energy between low energies and high energies is lost.  That is,
once the spectral dip is filled in, the low and high energy behaviors have
merged, and the sharp peak and broad hump at low temperature is simply
replaced by a single broad peak (with a leading edge gap due to the
pseudogap).  In fact, as a cautionary remark, it is obvious from Fig.~4b,
where the energy of the spectral peak is compared to the peak width, that
if anything, quasiparticles are at best marginally defined below 80K, and
become ill defined above this.  This is exactly the temperature at which
the weights plotted in Fig.~4a start to increase.  Still, everything is
consistent with the spectral peak simply losing its integrity as the
temperature is raised.  Certainly, the analysis here in no way supports a
picture of a well defined quasiparticle peak whose weight simply disappears
upon heating.  An analysis was also done with a background subtraction, and
similar conclusions were found.

Our results have significant implications for microscopic theories of
the cuprates.  In previous work \cite{NORM97,NDING,HUMP99}, we have
argued that the spectral lineshape at $(\pi,0)$
is naturally explained by the coupling of the electrons to a magnetic
resonance seen by neutron scattering.  Since the intensity of
this resonance decreases with temperature, then the coupling
of the electrons to this mode also decreases, and thus one would expect
$Z$ to decrease, just as we find in Fig.~3a.  As the neutron resonance
intensity decreases, the spin gap in the dynamic susceptibility fills in, which
we speculate is responsible for the ``filling in" of $Im\Sigma$ seen in
Fig.~2a.  The combination of these two effects causes the spectral
peak to rapidly broaden with temperature, losing its integrity above $T_c$.

In conclusion, we have studied the thermal evolution of the electron
self-energy by direct analysis of ARPES data, and come to the conclusion that
the spectral peak below $T_c$ broadens out above $T_c$ due to a ``lifetime
catastrophe", rather than the quasiparticle weight $z$ monotonically decreasing
to zero.

We thank Mohit Randeria for discussions.  This work was
supported by the U. S.  Dept. of Energy,
Basic Energy Sciences, under contract W-31-109-ENG-38, and the National 
Science Foundation DMR 9974401.

\begin{figure}
\epsfxsize=2.4in
\epsfbox{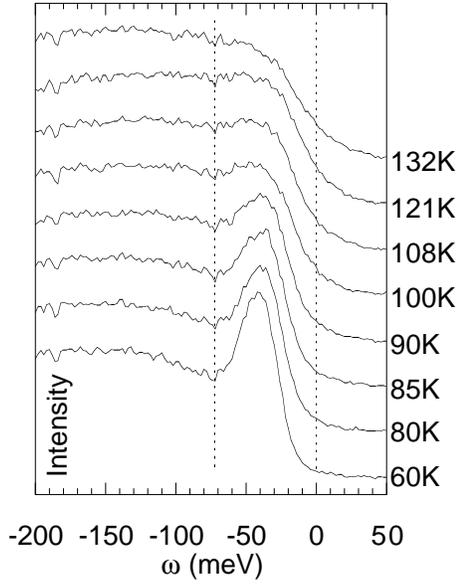}
\vspace{0.5cm}
\caption{Temperature dependence of ARPES data at $(\pi,0)$
for a $T_c$=90K Bi2212 sample.  The vertical dotted lines
mark the spectral dip energy and the chemical potential.}
\label{fig1}
\end{figure}

\begin{figure}
\epsfxsize=3.4in
\epsfbox{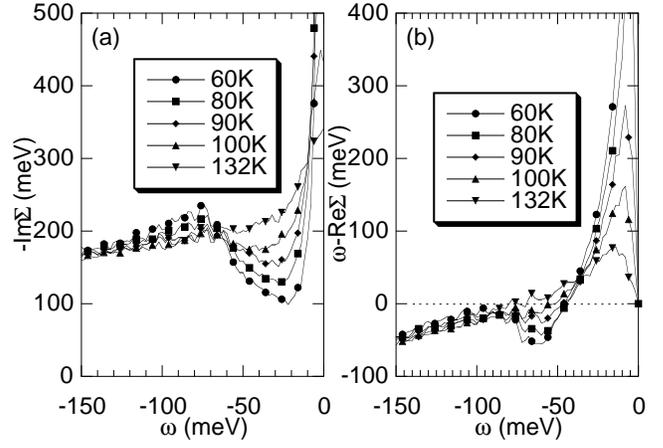}
\vspace{0.5cm}
\caption{Temperature dependence of (a) the imaginary and (b) real parts of the
self-energy derived from the data of Fig.~1.}
\label{fig2}
\end{figure}

\begin{figure}
\epsfxsize=3.4in
\epsfbox{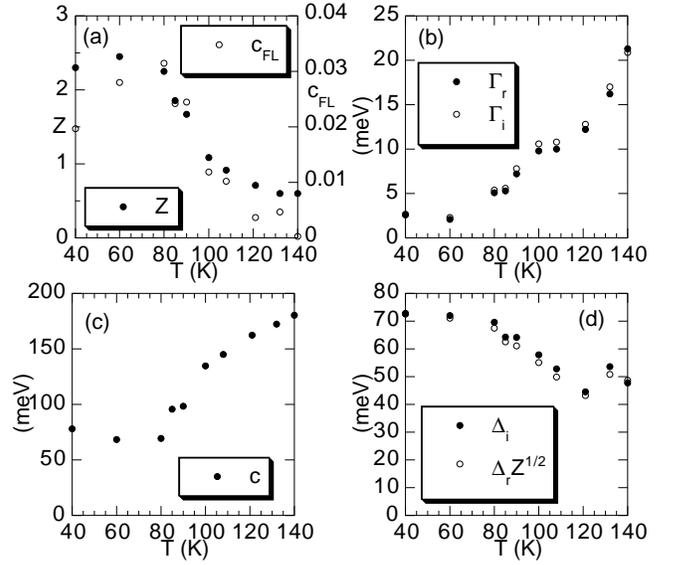}
\vspace{0.5cm}
\caption{Temperature dependence of the various coefficients of the self-energy
from Eq.~3 obtained from fits of Fig.~2 over an energy range of $\pm$60 meV.}
\label{fig3}
\end{figure}

\begin{figure}
\epsfxsize=3.4in
\epsfbox{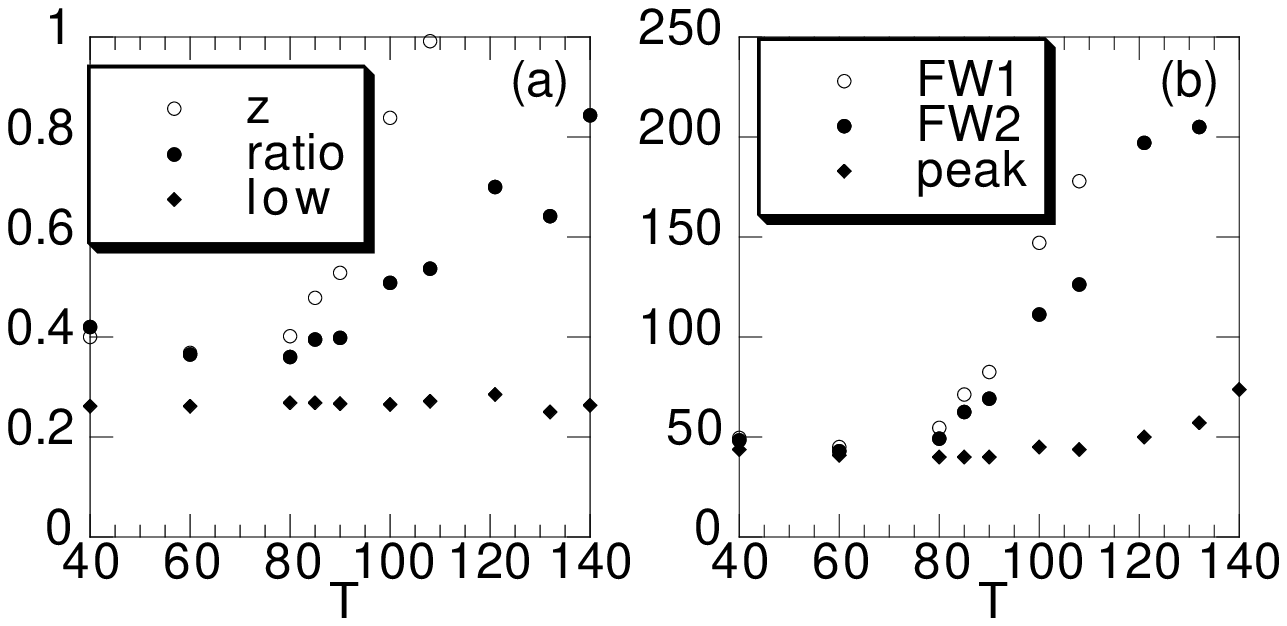}
\vspace{0.5cm}
\caption{Temperature dependence of (a) the spectral peak weight ($z$ and
$ratio$) and (b)
the width (FWHM) of the spectral peak ($FW1$ and $FW2$) as defined
in the text.  Also plotted is the low energy weight ($low$) and the peak
energy ($peak$).
For $FW2$, the
low energy edge of the peak is either defined by the HWHM, or zero energy
if $A$ at zero energy is greater than half the maximum.}
\label{fig4}
\end{figure}

\end{document}